\documentclass[%
 amsmath,amssymb,
 prfluids,
]{revtex4-2}

\usepackage{newtxtext}
\usepackage{newtxmath}
\usepackage{natbib}
\usepackage{graphicx}
\usepackage{amsmath}
\usepackage{amssymb}
\usepackage{soul}
\usepackage{caption}
\usepackage{natbib}
\usepackage{soul}
\usepackage{color}
\usepackage{hyperref}
\hypersetup{
    colorlinks = true,
    urlcolor   = blue,
    citecolor  = black,
}
\newcommand{\RomanNumeralCaps}[1]

\definecolor{burgundy}{rgb}{0.5, 0.0, 0.13}

\definecolor{past}{rgb}{0.47,0.87,0.47} 

\def\bea{\begin{equation}}
\def\eea{\end{equation}}

\newcommand{\pdiff}[2]{\frac{\partial #1}{\partial #2}}

\begin{document}

\preprint{APS/123-QED}

\title{Towards eliminating the nonlinear Kelvin wake}

\author{J. S. Keeler}
\email{j.keeler@uea.ac.uk}
\affiliation{School of Engineering, Mathematics and Physics, University of East Anglia, Norwich, NR4 7TJ, UK}
\author{B. J. Binder}
\email{benjamin.binder@adelaide.edu.au}
\affiliation{School of Computer and Mathematical Sciences, University of Adelaide, Australia}
\author{M. G. Blyth}
\email{m.blyth@uea.ac.uk}
\affiliation{School of Engineering, Mathematics and Physics, University of East Anglia, Norwich, NR4 7TJ, UK}

\begin{abstract}

The nonlinear disturbance caused by a localised forcing moving at constant speed on the free surface of a liquid of finite depth is investigated using the forced Kadometsev-Petviashvili equation. 
The presence of a steady v-shaped Kelvin wave pattern downstream of the forcing is established for this model equation, and the wedge angle is characterised as a function of the Froude number. Inspired by this analysis, it is shown that the wake can be eliminated via a careful choice of the forcing distribution and that, significantly, the corresponding nonlinear wave-free solution is stable so that it could potentially be seen in a physical experiment. The stability is demonstrated via the numerical solution of an initial value problem for which the steady wave-free state is attained in the long-time limit.

\end{abstract}
\maketitle
\section{Introduction}\label{sec:intro}

The study of the wake produced by a moving body on the surface of water is a classical problem in fluid dynamics. The v-shaped wake pattern that is seen behind such a body is a rare example of a fluid dynamics phenomenon that is well-known to everyone. Attempting to minimise or even eliminate the Kelvin wake presents an important problem since in many applications its presence leads to undesirable consequences. For example it can cause erosion to waterway and river banks \citep[e.g.][]{bishop}, and the wave drag it produces limits the fuel efficiency of boats and ships \citep{havelock3}. 

The mathematical description of the three-dimensional Kelvin wake has been well known for over a century \citep{kelvin1887ship}. 
Typically, a moving body such as a ship is represented by a pressure disturbance that propagates over the fluid surface. 
The angle subtended by the characteristic v-shape bounding the main surface 
disturbance is usually referred to as the Kelvin wedge angle, and is herein denoted by 
$\theta_{\mathrm{k}}$.
In deep water $\theta_{\mathrm{k}}$ can be found using a linear simplification of the full Euler system on the assumption that the disturbance to the surface is small. This leads to the result that $\tan^2\theta_k = 1/8$ yielding $\theta_{\mathrm{k}}\approx 19.47^\circ$ \citep[see, for example][]{whitham}. In shallow water $\theta_{\mathrm{k}}$ depends on the flow-speed and water depth 
\citep{rabaud2013ship}. 

Despite the ubiquity of the Kelvin wake phenomenon, and the large number of research articles that have been devoted to it, it remains a topic of considerable scientific interest. We highlight in particular two research areas of recent interest: (i) the numerical calculation of the fully nonlinear Kelvin wake and the associated wave-drag, and (ii) the theoretical description of Kelvin wakes for slow-moving vessels. For (i) the so-called `wave-drag coefficient' \citep{havelock3,pedersen1988three,li2016ship} can be used to quantify the resistance induced by the wake.
Recently, \cite{pethiyagoda2014apparent,buttle2018three,ctugulan2022three} used a numerical boundary-integral approach coupled with Newton-Krylov and pre-conditioning techniques to solve
the fully nonlinear Euler system for inviscid, irrotational flow (hereinafter referred to as the Euler system)
in order to capture efficiently the steady wave patterns behind a moving object, and provided 
important insight into the nonlinear wave-drag. 
Significantly, \cite{pethiyagoda2014apparent,buttle2018three} showed that as the flow-speed and forcing pressure amplitude are increased, nonlinearity, via the amplitude of the external forcing, alters the Kelvin wedge angle and therefore the physical extent of the wavy region. 
In the aforementioned studies, the authors made simple choices for the external forcing functions, for example using combinations of Gaussian forcing functions. There has also been work done to quantify the wave-drag for an arbitrary forcing distribution using the linearised form of the Euler system \citep{miao2015wave}. In none of these studies did the wave-pattern completely disappear for any choice of the external forcing. 

Regarding (ii), for slow-moving boats the Froude number, $Fr = {U}/\sqrt{gH}$, is small, where $U$, $g$ and $H$ are a typical flow speed, the gravitational acceleration, and a typical length scale, respectively. Assuming $Fr\ll 1$, an asymptotic solution of the two-dimensional Euler system that utilises expansions in powers of $Fr$, somewhat surprisingly, produces no waves at 
 each order of $Fr$. This apparent paradox can be resolved by resorting to `beyond-all-orders' asymptotics, and in this context it can be shown that there are waves present downstream that are exponentially small in $Fr$ \citep{trinh2011waveless,trinh2014wake}. For the 
 two-dimensional problem, \cite{trinh2011waveless} showed that downstream waves cannot be eliminated for a single-cornered ship. However, \citep{trinh2014wake} were not able to make such a conclusive statement for more general piecewise-linear hulls and they did not discount the possibility of wave-free solutions. 
\cite{lustri2013steady} and \cite{pethiyagoda2021kelvin} considered the linear three-dimensional problem and characterised the wedge angle in terms of the system parameters. Recently, as a first step towards examining the Euler system, \cite{kataoka2023nonlinear} showed how an exponential asymptotics approach may be used for smooth boat hulls in the context of a model equation. Examining smooth hulls is important as they can be used to model the bulbous hull shapes that are known to minimise wave-drag \cite[see, for example][]{grosenbaugh1989nonlinear}. However, modelling smooth bodies moving at arbitrary speeds using the Euler system remains an open challenge.

In this paper, we work towards the goal of eliminating the Kelvin wake behind a moving body in a real flow by studying the simplified model 
Kadometsev-Petviashvili equation with a localised forcing term. This equation was originally derived by \cite{katsis1987excitation}, and it is hereinafter referred to as fKP. 
We demonstrate that for this equation a generic forcing produces a trailing wave pattern that is similar to that found for the Euler system with a v-shaped wake whose angle we characterise as a function of the Froude number. This includes the critical case $Fr=1$ covered by  \cite{katsis1987excitation}, who showed that under this condition the wake fills the entire region behind the disturbance.
Furthermore, we use a simple mathematical argument to show that for the fKP system it is  possible to eliminate the wake. By a judicious choice of forcing function, we construct solutions for which the disturbance to the water surface is localised around the forcing so that the surface is undisturbed in all directions into the far field. We refer to such solutions as \textit{wave-free} solutions. Strikingly, we provide evidence that the wave-free solutions are stable, meaning that they are reached in the long-time limit of an initial value problem (IVP), leading to the possibility of observing them in a physical experiment.

We proceed as follows. In \S~\ref{sec:kelvinwaves} we describe the problem statement including the fKP equation and boundary conditions, and we characterise the v-shaped Kelvin wake-pattern that emerges for a generic choice of forcing function. In \S~\ref{sec:wave_free} we discuss how to construct a wave-free steady state for the fully nonlinear fKP equation in the presence of a suitably chosen localised forcing. In \S~\ref{sec:stability} we formulate and solve numerically an IVP to show that the wave-free states are stable. Finally, in \S~\ref{sec:conclusion} we present our conclusions and highlight future directions for study.

\section{Steady Kelvin wake-patterns of the fKP}\label{sec:kelvinwaves}

Before we demonstrate how we construct wave-free solutions, it is instructive to characterise the v-shaped Kelvin wake-pattern, and in particular the wedge angle, for a generically chosen external forcing in terms of the flow-speed. The procedure we follow is well known for the Euler system \citep[see, for example][]{miao2015wave} and has recently been reconstructed for a different model PDE (partial differential equation) by \cite{kataoka2023nonlinear}.

As derived in the context of water-waves the fKP is a PDE for the height of the free-surface, denoted $\eta(\boldsymbol{x},t)$, in terms of an external forcing profile, $\sigma(\boldsymbol{x})$ (either representing a surface pressure distribution or the shape of a topographic obstacle), where $\boldsymbol{x} = [x,y]^T$ and the prevailing flow is in the $x$ direction \citep{katsis1987excitation}. It is derived from the Euler system under the assumptions of i) shallow-water, ii) small forcing amplitude, iii) weak $y$-dependence and iv) $Fr\approx 1$ \citep[see, for example][for mathematically rigorous statements on the validity of this model]{lannes}. Therefore, in constrast to the small-$Fr$ literature stated in the introduction, our focus is on moderate and fast-moving boats/topographic obstacles. The equation is 
\bea
\left(\eta_t + (Fr - 1)\eta_x - \frac{3}{2}\eta\eta_x - \frac{1}{6}\eta_{xxx} - \frac{1}{2}\sigma_x\right)_x-\frac{1}{2}\eta_{yy} = 0,\qquad x,y\in\mathbb{R},
\label{timeforcedKP}
\eea
where the subscripts indicate partial derivatives. Far upstream of the forcing, the free-surface is assumed flat and hence we impose the boundary conditions
\begin{align}
  &\eta,\eta_x,\eta_{xx},\eta_{xxx}\to 0,\qquad \mbox{as} \qquad x\to-\infty, \label{inflowbc}\\
  &\eta_y \to 0,\qquad\mbox{as}\qquad y \to \pm \infty.\label{sidebc}
\end{align}
We note that \eqref{timeforcedKP} is valid in the context of no/weak surface tension, and is a modification of the KPII equation (the KPI equation has the opposite sign on the fourth derivative and is for strong surface tension). We emphasise that we were unable to find any explicit analysis for the Kelvin wedge angle in \eqref{timeforcedKP} in the literature, but note the linear analysis of \cite{katsis1987excitation}; restricted to critical flow, $Fr=1$, and a time-dependent response with no explicit determination of the Kelvin wedge angle as a function of $Fr$.

To determine the Kelvin wedge angle, we examine linear steady solutions to \eqref{timeforcedKP} by assuming $\eta$ is small;
\bea
\mathcal{L}(\eta_{\ell})= 3\sigma_{\ell,xx},\qquad \mathcal{L} \equiv 6(Fr-1)\partial_{xx} - \partial_{xxxx} - 3\partial_{yy} = 0.
\label{steadyfkp}
\eea
For convenience, in what follows, $\eta_{\ell},\sigma_{\ell}$ correspond to a linear solution and forcing term of \eqref{steadyfkp}, while $\eta,\sigma$ correspond to a nonlinear solution and forcing term of \eqref{timeforcedKP}.

We solve \eqref{steadyfkp} using the two-dimensional Fourier transform, defined as
\bea
\hat{\eta}_{\ell}(\boldsymbol{k}) = \mathcal{F}(\eta_{\ell}) \equiv \frac{1}{2\pi}\int_{-\infty}^{\infty}\int_{-\infty}^{\infty}\eta_{\ell}(\boldsymbol{x})\mbox{exp}(-\mathrm{i}\boldsymbol{k}\cdot\boldsymbol{x})\,\mbox{d}x\,\mbox{d}y,
\eea
where $\boldsymbol{k} = [k_x,k_y]^T$. Introducing a polar coordinate system;
\bea
\boldsymbol{x} = r[\cos\theta,\sin\theta]^T,\qquad \boldsymbol{k} = \kappa[\cos\varphi,\sin\varphi]^T
\eea
and letting $\hat{\sigma}_{\ell}(\boldsymbol{k}) = \mathcal{F}(\sigma)$, the steady linear solution to \eqref{steadyfkp} can be written in terms of the inverse Fourier transform as
\bea
\eta_{\ell}(\boldsymbol{x}) = \frac{1}{2\pi}\int_{0}^{2\pi}\int_{0}^{\infty}\frac{3\hat{\sigma}_{\ell}(\boldsymbol{k})\,\kappa\,\mbox{exp}(\mathrm{i}\kappa r\cos(\varphi-\theta))}{(\kappa^2-\kappa_0^2)\cos^2\varphi}\,\mbox{d}\kappa\,\mbox{d}\varphi,\quad\kappa_0^2 = \frac{3\tan^2\varphi - 6(Fr-1)}{\cos^2\varphi}.
\label{linearsol}
\eea
The Kelvin wedge angle can be determined by examining the leading-order asymptotic behaviour of \eqref{linearsol} far downstream, as $r\to\infty$, $|\theta|<\pi/2$. To proceed, the inner $\kappa$-integration can be achieved by extending the domain of integration into the complex plane, i.e. $\kappa\in\mathbb{C}$, and then using Cauchy's residue theorem. For subcritical and critical flow, $Fr<1$ and $Fr=1$, respectively, $\kappa_0^2>0$, and hence there are simple poles at $\kappa=\pm\kappa_0$; we only consider the pole at $\kappa=+\kappa_0$ as the integration limits in \eqref{linearsol} is the positive real line. If $Fr>1$, it is possible that $\kappa_0^2<0$ and this will be discussed below.

Great care has to be taken when choosing the path of integration and choice of contour indentation around the pole $\kappa=+\kappa_0$. For $\cos(\varphi-\theta)>0$, we close our path of integration in the upper half complex $\kappa-$plane by adding a quarter-circle arc from the positive real $\kappa$ axis to the positive imaginary $\kappa$ axis and then returning back to the origin (we note that the contribution of the integrand on the imaginary axis is exponentially small in $r$). The boundary conditions, \eqref{inflowbc}, dictate there are no waves far upstream and hence if $\cos\theta<0$ ($x<0$) we indent the contour on the real $\kappa$-axis so it goes over the pole and hence make no contribution to the integral. Far downstream, when $\cos\theta > 0$, when waves can appear due to the radiation condition, the contour is indented under the pole; therefore making a non-zero contribution due to the residue of the integrand at $\kappa_0$. For supercritical flow, $Fr>1$, $\kappa_0^2$ will become negative when $\tan^2\varphi<2(Fr-1)$ and the pole will be located on the imaginary axis, but in this case the residue is now exponentially small as $r\to\infty$. We note that for $\cos(\varphi-\theta)<0$ the contour gets deformed onto the negative imaginary axis and the preceding description of the contour is reversed.

Assuming $\cos\theta>0$ and the pole is on the real axis if $Fr>1$, by calculating the residue of the pole, the leading-order asymptotic behaviour far downstream is 

\bea
\eta_{\ell}(\boldsymbol{x}) \sim \frac{3}{2}\mathrm{i}\int_{-\pi/2+\theta}^{\pi/2}\frac{\hat{\sigma}_{\mathrm{\ell}}(\boldsymbol{k}_0)}{\cos^2\varphi}\mbox{exp}(\mathrm{i}|\kappa_0| r\cos(\varphi-\theta))\,\mbox{d}\varphi,\qquad \mbox{as}\qquad r\to\infty,
\label{varphiint}
\eea
where $\boldsymbol{k}_0 = |\kappa_0|[\cos\varphi,\sin\varphi]^T$ is dependent on $\varphi$ only. The limits of integration on the outer-$\varphi$ integral have changed from \eqref{linearsol} as the solution is symmetric about $\theta=0$ so we only need to consider $0<\theta<\pi$, which means that as $\cos(\varphi-\theta)>0$, $-\pi/2+\theta<\varphi<\pi/2$ (if $\cos(\varphi-\theta)<0$ then $-\pi/2<\varphi<\pi/2-\theta$). The dominant behaviour of the integral as $r\to\infty$ occurs when the exponent in \eqref{varphiint} is stationary, i.e.
\bea
\pdiff{g}{\varphi} = 0, \: \mbox{where} \: g(\varphi,\theta) = \kappa_0\cos(\varphi-\theta)\equiv \frac{\sqrt{3}(\tan^2\varphi - 2(Fr-1))^{1/2}\cos(\varphi-\theta)}{\cos\varphi},
\label{kelvinequation}
\eea
The leading-order asymptotic behaviour of \eqref{varphiint} as $r\to\infty$ is readily obtained using the stationary-phase formula 
\bea
\eta_{\ell}(\boldsymbol{x})\sim \frac{3\sqrt{\pi}(\mathrm{i}-\lambda)}{4}\sum_{n}r^{-1/2}\,\frac{\hat{\sigma}_{\mathrm{\ell}}(\kappa_0,\varphi_n)}{|g''(\varphi_n)|^{1/2}\cos^2\varphi_n}\mbox{exp}\left[\mathrm{i}|\kappa_0| r\cos(\varphi_n-\theta) \right]+ \mbox{c.c},
\label{stat_phase}
\eea
where the sum is over the $n$ permissible solutions, $\varphi=\varphi_n$, of \eqref{kelvinequation}, $\lambda = \mbox{sgn}(g''(\varphi_n))$ is the signum function, $g''(\varphi_n)$ is the second derivative of $g(\varphi)$ with respect to $\varphi$, $\hat{\sigma}_{\ell}(\kappa_0,\varphi_n) = \hat{\sigma}_{\ell}(\boldsymbol{k}_0)|_{\varphi = \varphi_n}$ and c.c stands for complex conjugate. Physically, as can be seen from \eqref{stat_phase}, $\kappa_0$ characterises the wave number of the oscillations far downstream. 

\begin{figure}
  \centering
  \includegraphics[width=\textwidth,trim= 0 70 0 0,clip]{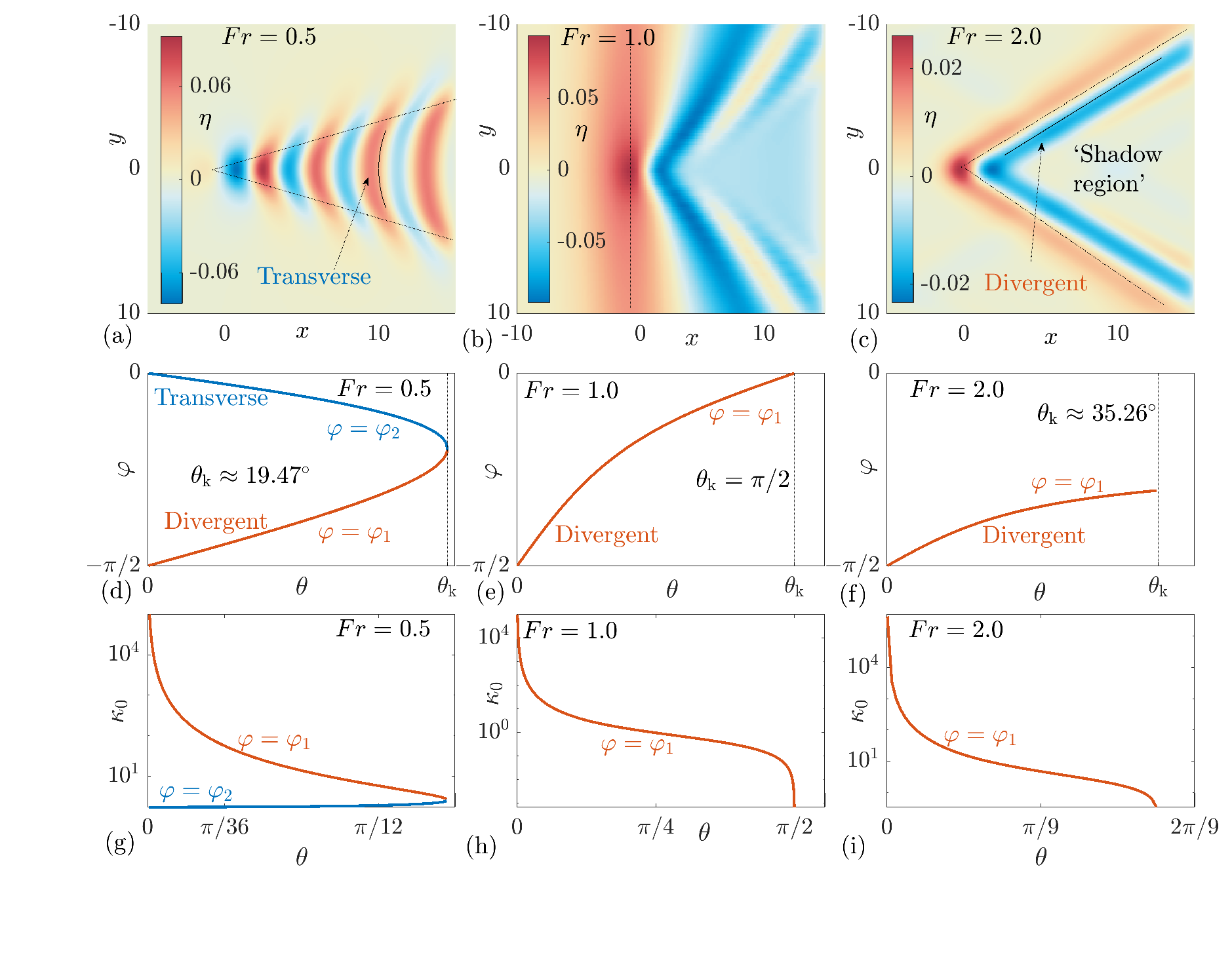}
  \caption{The left, middle and right columns show results for $Fr=0.5,1.0$ and $2.0$, respectively. (a)-(c) Nonlinear steady solutions of \eqref{timeforcedKP} for $\sigma(\boldsymbol{x}) = a\,\mbox{exp}(-\boldsymbol{x}\cdot\boldsymbol{x})$ with $a=0.1$. The linear Kelvin wedge angle, $\theta_{\mathrm{k}}$ is indicated as dotted lines. (d)-(f) The roots of \eqref{kelvinequation} in the range $-\pi/2 + \theta < \varphi <\pi/2$ as a function of $\theta$ for a general forcing function $\sigma_{\ell}$. (g)-(i) The pole on the real axis, $\kappa_0$, evaluated at $\varphi_{1,2}$.}
  \label{fig:angle_solutions}
\end{figure}

When $Fr\neq 1$ the solutions satisfy the quadratic equation for $\tan\varphi$,
\bea \label{Fgt1eq}
2\sin \theta \tan^2\varphi + \cos \theta \tan \varphi - 2(Fr-1) \sin \theta = 0.
\eea
Hence when $Fr<1$ there are two real solutions for $\varphi$ provided that
\bea
0<|\theta|<\theta_{\mathrm{k}}(Fr),\qquad \mbox{where}\qquad 16(1 - Fr)\tan^2\theta_{\mathrm{k}} = 1.
\label{wedge}
\eea
The value of $\theta_{\mathrm{k}}(Fr)$ is interpreted as the Kelvin wedge angle. It is indicated by the dashed vertical line in figure~\ref{fig:angle_solutions}(d) for the case $Fr=0.5$. Comparing the value of $\kappa_0$ at each root, see figure~\ref{fig:angle_solutions}(g), we see that one of the solutions, $\varphi_1$, is short-wavelength and corresponds to a so-called \textit{divergent} wave (that shall be discussed below) whist the other solution, $\varphi_2$, is long-wavelength and corresponds to a so-called \textit{transverse} wave that oscillates on the line $\theta=0$ as can be seen in \eqref{stat_phase}. For historical context, the divergent/transverse terminology was first coined in \cite{havelock0}. We note the special case of $Fr=1/2$, as shown in figure~\ref{fig:angle_solutions}(d),(g), which results in $\theta_{\mathrm{k}} = \mathrm{arctan}(1/2\sqrt{2})\approx 19.47^{\circ}$, the classical Kelvin wedge angle for the linearised Euler system. Interestingly, this is the value of $Fr$ reported by \cite{rabaud2013ship} as approximately the transition point between a Kelvin wake-regime and a Mach wake-regime, where all linear waves travel with equal speed.  
\begin{figure}
  \centering
  \includegraphics[width=0.62\textwidth,trim= 0 0 0 0,clip]{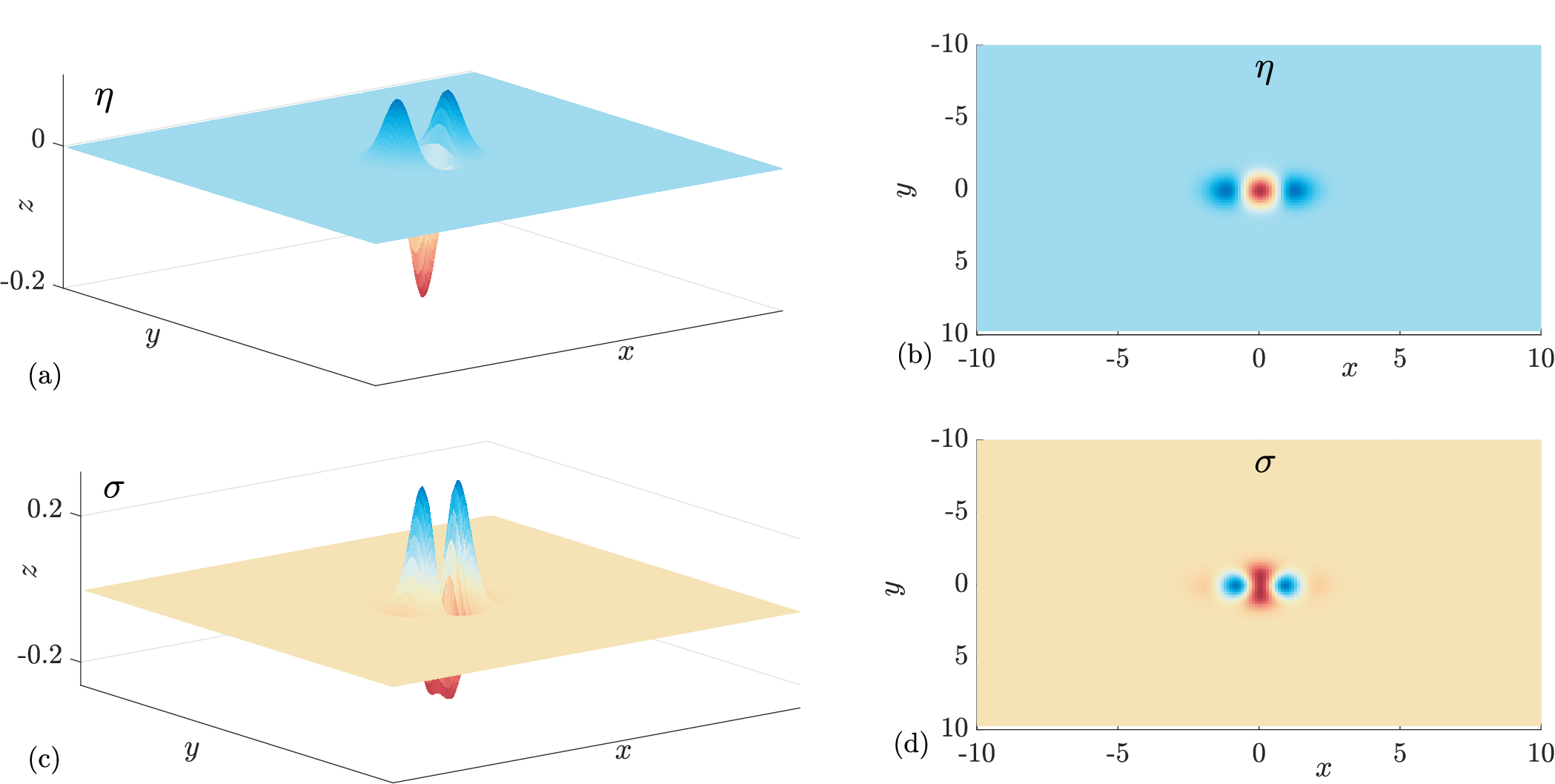}
  \caption{Kelvin wedge angle dependence on $Fr$ using \eqref{wedge} and \eqref{realroot}.}
  \label{fig:wedge_angle_solution}
\end{figure}

When $Fr>1$ equation \eqref{Fgt1eq} has two solutions for any $0< |\theta| < \pi/2$. However, for one of these solutions, $\varphi_2$, we find that $\kappa_0^2<0$ for all $0< |\theta| < \pi/2$ and so the pole in \eqref{linearsol} is not on the real axis and does not contribute to the far-field wave pattern. For the other (divergent) solution, $\varphi_1$, the transition from $\kappa_0^2>0$ (pole on the real axis) to $\kappa_0^2<0$ (pole on the imaginary axis) occurs when 
$\tan^2\varphi=2(Fr-1)$. Inserting the latter into \eqref{Fgt1eq} we find that the permissible $\theta$ are bounded by the Kelvin wedge angle, $\theta_{\mathrm{k}}$ and satisfies
\bea
0<|\theta|<\theta_{\mathrm{k}}(Fr),\qquad\mbox{where}\qquad 2(Fr-1)\tan^2\theta_{\mathrm{k}} = 1.
\label{realroot}
\eea

Focussing on $Fr>1$, a key feature of the divergent wave solution is the presence of a `shadow-region' downstream of the forcing where there is minimal surface disturbance, see figure~\ref{fig:angle_solutions}(c), and no surface disturbance at $\theta=0$. To see this, $\varphi_1\to-\pi/2$ and $\kappa_0\to\infty$ as $\theta\to 0$, see figure~\ref{fig:angle_solutions}(f)(i). Therefore, examining \eqref{stat_phase}, for a fixed value of $r$; $\eta_{\ell}$ decays to zero as $\theta\to 0$, provided $\hat{\sigma}_{\ell}(\kappa_0,\varphi_1)$ decays sufficiently quickly as $\kappa_0\to\infty$. The Gaussian forcing we choose in the calculations in figure~\ref{fig:angle_solutions}(a)-(c), and discuss below, fulfills this criteria. 

Finally, when $Fr=1$ \eqref{kelvinequation} has a unique solution, $\varphi_1$, corresponding to a divergent wave, for any $0<|\theta|\leq\pi/2$ which satisfies
\bea\label{F=1eq}
\tan \varphi = -\frac{1}{2}\cot \theta.
\eea
Hence the Kelvin wedge angle is $\theta_{\mathrm{k}} = \pi/2$, as was first noted in \cite{katsis1987excitation}.

In summary, we have determined the far-field Kelvin wedge angle as a function of $Fr$ for subcritical and supercritical flow, as stated in \eqref{wedge} and \eqref{realroot}, respectively. The wedge angle is plotted against $Fr$ in figure~\ref{fig:wedge_angle_solution}.
We note that $\theta_{\mathrm{k}}\to\pi/2$ as $Fr\to 1^{\pm}$ and that for large $Fr$ we have the asymptotic behaviour 
\bea
\theta_{\mathrm{k}}\sim \, (\sqrt{2}/2)\,Fr^{-1/2}\qquad\mbox{as}\qquad Fr\to \infty.\label{largebehaviour}
\eea 
For the Euler system and linearised Euler system, the dependence of the flow-speed on $\theta_{\mathrm{k}}$ has been reported before, where it was found that increasing the forcing amplitude has the effect of increasing the wake angle beyond the classical Kelvin value \citep{pethiyagoda2014apparent}, but increasing the flow-speed has the effect of decreasing the wake angle \citep{rabaud2013ship}. The leading-order asymptotic behaviour in \eqref{largebehaviour} differs to the $\theta_{\mathrm{k}}\sim \,(1/2\sqrt{2\pi})\,Fr^{-1}$ predicted by \cite{rabaud2013ship,miao2015wave}. This discrepancy can be explained by the fact that the fKP is only valid around $Fr\approx 1$.

We can compare the values obtained from the solution to \eqref{wedge} and \eqref{realroot} with the numerically computed nonlinear steady wave-patterns, found by solving the steady form of \eqref{timeforcedKP}, as shown in figure~\ref{fig:angle_solutions}(a)-(c) for $Fr=0.5,1.0,2.0$, respectively. For concreteness, we choose a Gaussian forcing $\sigma(\boldsymbol{x}) = a\,\mbox{exp}(-\boldsymbol{x}\cdot\boldsymbol{x})$ and $a=0.1$. The dashed lines in each panel indicate the Kelvin wedge angle, $\theta_{\mathrm{k}}$. This theoretically derived angle shows excellent agreement with the numerically produced nonlinear wave patterns, giving us confidence in our numerical scheme (based on a variational formulation of \eqref{timeforcedKP} and discretised using finite-elements). Furthermore, the wave-patterns are qualitatively the same as the fully nonlinear solutions reported in \cite{buttle2018three}, where transverse waves are present for $Fr<1$ but absent when $Fr\geq1$, where instead a `shadow-region' appears. 

To our knowledge this is the first time the Kelvin wedge angle and wave-pattern has been calculated explicitly for a range of $Fr$, in the context of the fKP equation and is the first main result of this paper. Due to the simplicity of \eqref{steadyfkp}, this demonstrates the promising potential of the fKP equation as a `sand-pit' model to understand nonlinear Kelvin wave-patterns.

An important part of the analysis that we emphasise, is the importance of the pole, $\kappa_0$, in producing waves downstream; if we can eliminate this pole then there is a possibility a wave-free steady state could exist. Since $\kappa_0$ does not depend on the explicit form of the forcing term, $\sigma(\boldsymbol{x})$, then the forcing does not play a leading role in determining the wave-pattern far downstream. It is often convenient to model $\sigma(\boldsymbol{x})$, taking the role as a `boat', as a Dirac-delta function or Gaussian for a one-point wavemaker \citep[see, for example][]{whitham,katsis1987excitation} or as a dipole modelling a two-point wavemaker \citep{noblesse2014can,miao2015wave}. We will now show that through a judicious choice of ${\sigma}(\boldsymbol{x})$, corresponding to a localised forcing term, a nonlinear steady wave-free $\eta(\boldsymbol{x})$ exists.

\section{Nonlinear wave-free steady solutions}\label{sec:wave_free}

Initially, we concentrate on the linear solution \eqref{linearsol}, which in Cartesian coordinates ($\boldsymbol{k} = [k_x,k_y]^T$) can be written as
\bea
\eta_{\ell}(\boldsymbol{x}) =\frac{1}{2\pi} \int_{-\infty}^{\infty}\int_{-\infty}^{\infty}\frac{3k_x^2\,\hat{\sigma}_{\ell}(\boldsymbol{k})}{6(Fr-1)k_x^2 + k_x^4 - 3k_y^2}\,\mbox{exp}(\mathrm{i}\boldsymbol{k}\cdot\boldsymbol{x})\,\mbox{d}k_x\,\mbox{d}k_y.
\label{etafft}
\eea
As previously stated, the pole in the integrand is responsible for the far-field wave pattern as $x\to\infty$. Therefore, an obvious choice for $\hat{\sigma}_{\ell}(\boldsymbol{k})$ that eliminates this pole is
\bea
\hat{\sigma}_{\ell}(\boldsymbol{k}) = \frac{1}{3}[6(Fr-1)k_x^2 + k_x^4 - 3k_y^2]\,\hat{f}(\boldsymbol{k}),
\label{sigmafft}
\eea
where $\hat{f}$ is the Fourier transform of an arbitrary function, $f(\boldsymbol{x})$, that is localised in physical space. Therefore, in physical space, the linear solution is 
\bea
\eta_{\ell}(\boldsymbol{x}) = \frac{1}{2\pi}\int_{-\infty}^{\infty}\int_{-\infty}^{\infty} k_x^2 \hat{f}(\boldsymbol{k})\,\mbox{exp}(\mathrm{i}\boldsymbol{k}\cdot\boldsymbol{x})\,\mbox{d}k_x\,\mbox{d}k_y = -(f(\boldsymbol{x}))_{xx}.
\label{wavefreelin}
\eea
A simple illustrative example is the choice 
$
\hat{f}(\boldsymbol{k}) =-a\,\pi\,\mbox{exp}\left(-(1/4)\boldsymbol{k}\cdot\boldsymbol{k}\right),\, a\in\mathbb{R}.
$
In this case it is straightforward to see from \eqref{wavefreelin} that
\bea
\eta_{\ell}(\boldsymbol{x}) = a\left(\mbox{exp}(-\boldsymbol{x}\cdot\boldsymbol{x})\right)_{xx},
\label{choice2}
\eea
which is, crucially, wave-free in the far-field; thus achieving our goal. In the results that follow we concentrate on the example highlighted above in \eqref{choice2}. Recall that $\eta_{\ell}$ and $\sigma_{\ell}$ satisfy \eqref{steadyfkp} so, in order to satisfy the nonlinear problem \eqref{timeforcedKP}, we choose 
\bea
\eta(\boldsymbol{x}) = \eta_{\ell}(\boldsymbol{x}),\qquad \sigma(\boldsymbol{x}) = \sigma_{\ell}(\boldsymbol{x}) - \frac{3}{2}\eta_{\mathrm{\ell}}^2(\boldsymbol{x}).
\label{choice3}
\eea

\begin{figure}
  \centering
  \includegraphics[width=1\textwidth,trim = 0 0 0 0,clip]{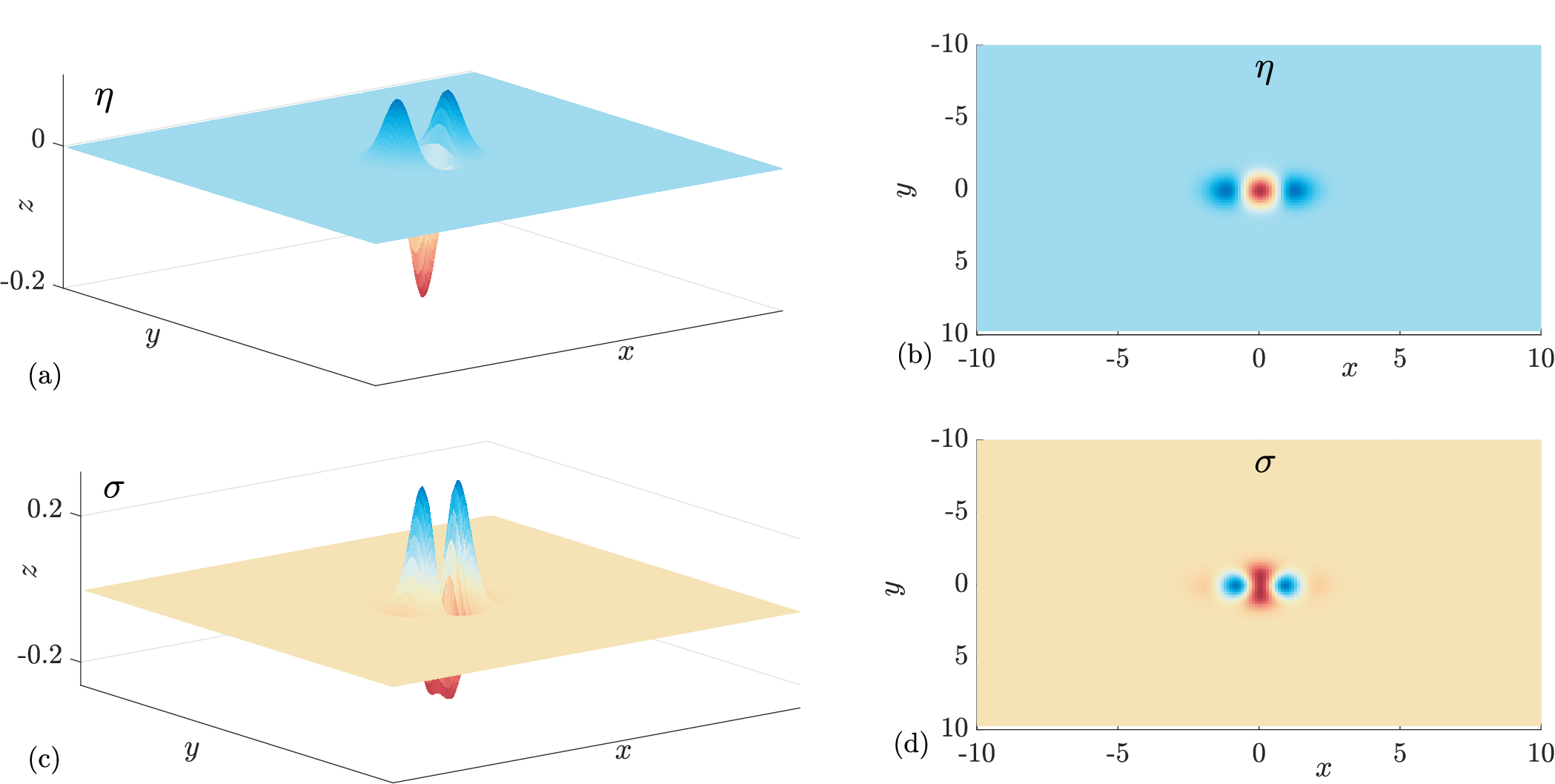}
	\caption{Steady nonlinear wave-free solutions defined in \eqref{choice3}. Panels (a)-(b); the nonlinear free-surface, $\eta(\boldsymbol{x})$ for $a=0.1,Fr=1.0$. Panels (c)-(d); the forcing, $\sigma(\boldsymbol{x})$.}
  \label{fig:steady_solutions}
\end{figure}
As an example, when $a=0.1,Fr=1.0$, the nonlinear solutions $\eta$ and $\sigma$ in \eqref{choice3} are shown in figure~\ref{fig:steady_solutions} panels (a),(b) and (c),(d), respectively. This is our second main result of the paper; we can choose a localised forcing function that results in a nonlinear wave-free, localised free-surface.

The forcing distribution in \eqref{choice3} is multi-signed which is similar to the dipole model of a monohull used by \cite{noblesse2014can} and for more arbitrary pressure shapes in \cite{miao2015wave}. Although it is not difficult to determine the solution in \eqref{choice3} without resorting to a Fourier analysis, the strategy of eliminating the pole in the linear system does hint at a possible approach for minimising the waves in the Euler system, where the nonlinearity is not as straightforward as the quadratic term in \eqref{timeforcedKP}. Finally we note that the steady free surface profile is sensitive to the choice of forcing in the sense that only a minor modification to \eqref{choice3} will lead to the reappearance of the Kelvin wake.

  \section{Stability of nonlinear wave-free steady solutions}\label{sec:stability}
  We now examine the solutions of \eqref{timeforcedKP} with \eqref{choice3} as the forcing term and \eqref{choice2} as the steady state. To probe the stability properties we solve a suite of IVPs numerically. We choose the initial condition
\begin{equation}
  \eta(\boldsymbol{x},t=0) = 0,
  \label{initial_condition}
\end{equation}
which represents a flat free-surface. To benchmark our results, initially we solve the IVP with a Gaussian forcing function, $\sigma(\boldsymbol{x}) = a\,\mbox{exp}(-\boldsymbol{x}\cdot\boldsymbol{x})$ and a dipole forcing $\sigma(\boldsymbol{x}) = a\,(\mbox{exp}(-\boldsymbol{x}\cdot\boldsymbol{x}))_x$ ($a=0.001$) and confirm that the system evolves towards a steady nonlinear Kelvin-wake, see \texttt{movie\_1.mp4} and \texttt{movie\_2.mp4} in the supplementary material.

Now, we concentrate on our special form of forcing in \eqref{choice2} and, to be consistent with the steady results in figure~\ref{fig:steady_solutions}, choose $Fr=1.0,a=0.1$. Figure~\ref{fig:inverse_subcrit} (corresponding to the animation \texttt{movie\_3.mp4} in the supplementary material) show the resulting time-dependent behaviour starting from \eqref{initial_condition}. As can be seen from the time-snapshots, for approximately $0<t<7.7$ (top row) a curved wave pattern emerges directly downstream of the forcing. Then, for approximately $t\geq 7.7$ (bottom row) this wake-pattern propagates downstream in the form of curved `ripples' before eventually separating and leaving the steady state, $\eta(\boldsymbol{x},t)$ which we have confirmed is the steady state in \eqref{choice3}. This behaviour is generic for a wide range of parameter space ($Fr\in[0.2,1.8]$ and $a\in[-0.1,0.1]$) and has been thoroughly checked for numerical convergence. This is strong numerical evidence that the wave-free steady state is not only stable but is also the asymptotic state of the system as $t\to\infty$. Furthermore, the waves emitted downstream in the initial stage of evolution appear different to the classical Kelvin wake-pattern and these `ripples' may be linked to the recent work of \cite{zhang2023rare}, where ripple-solutions were reported for the unforced KP equation. 
As well as the initial condition in \eqref{initial_condition}, the steady state is also stable to two- and three-dimensional perturbations as shown in \texttt{movie\_4.mp4} and \texttt{movie\_5.mp4}. Viewed altogether, these calculations show the remarkable stability of the wave-free steady state; our third, and perhaps most significant result of the paper. A carefully constructed forcing term, as described in the previous section, will result in a wave-free profile that can be observed in a physical experiment.

\begin{figure}
  \centering
  \includegraphics[width=1.1\textwidth,trim = 90 0 0 0,clip]{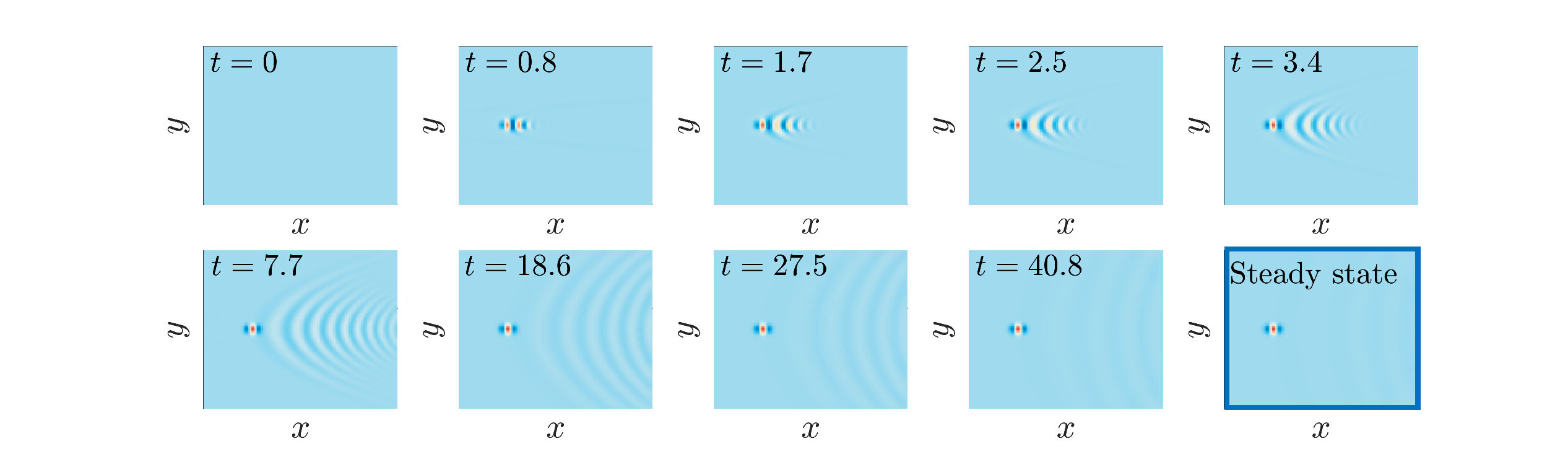}
	\caption{Time-dependent nonlinear solution, $x\in[-10,30],y\in[-20,20]$, starting from \eqref{initial_condition} with forcing in \eqref{choice2}, $Fr=1.0, a=0.1$.The corresponding animation is shown in \texttt{movie\_3.mp4} in the supplementary material.}
  \label{fig:inverse_subcrit}
\end{figure}

\section{Conclusions and perspective}\label{sec:conclusion}

We have demonstrated that the locally-forced Kadometsev-Petviashvili (fKP) equation is capable of producing the v-shaped Kelvin wave pattern that is observed in practice when an object moves along the surface of water, and we have characterised the associated wedge angle as a function of the Froude number. More importantly we have shown that, for the fKP, a judiciously chosen forcing can produce a steady wave-free surface, 
meaning that the disturbance caused by the forcing is confined to its neighbourhood so that the surface is flat in the far-field.
Crucially, using numerical simulations we have demonstrated that the wave-free states are stable in the sense that they are reached as $t\to\infty$ for a suitably posed IVP. 

Despite the simplicity of our mathematical argument, this appears to be the first time that wave-free steady solutions have been constructed for the fKP system and their stability properties 
calculated. Our results provide some evidence that it may be possible to eliminate the Kelvin wake in a real world setting. However, further work is needed to establish whether similar results can be achieved for other model systems or for the fully nonlinear Euler system, and 
whether wave-free solutions can be observed in a laboratory experiment. On the latter
point, we highlight the interesting recent work by \cite{euve2024asymetric} in which asymmetric 
wake patterns have been observed experimentally. 

Regarding applications, our results are at this point only 
suggestive. Nevertheless, they may ultimately pave the way to improving marine vessel design including the reduction of wave drag and the minimisation of the wake signature, as is desirable 
for `stealth' boats, for example. The localised forcing in our system may also 
be treated as a model for a topographic feature on the bottom, in which case our wave-free 
solutions suggest the possibility of designing underwater structures that minimise or eliminate any disruption to the water surface.

For the fKP problem studied here, the free-surface profile is extremely sensitive to small changes in the forcing function, and we therefore anticipate that the forcing that produces wave-free solutions for fKP may not work so well for other model equations (e.g. the Green-Naghdi equations) or for the full Euler system. 
While various pressure distributions have been investigated for the linearised Euler system \citep{miao2015wave}, the fully nonlinear Euler system presents a considerable challenge, and is the subject of our ongoing work.

\vskip2pc


\bibliographystyle{jfm}
\bibliography{water_wave}

\end{document}